\documentclass[aps,pre,preprint,superscriptaddress]{revtex4-1}

\usepackage{bm}
\usepackage{amsmath}
\usepackage[dvipdfmx]{graphicx}
\usepackage{setspace}
\usepackage{color}
\usepackage{lineno}
\begin{document}

%%%% Article title to be placed here
\title{Reinforcement learning account of network reciprocity}

\author{Takahiro Ezaki}
\affiliation{PRESTO, Japan Science and Technology Agency, 4-1-8 Honcho, Kawaguchi, Saitama, Japan}
%%%%%%%%% Insert author address here
%%%% Subject entries to be placed here %%%%
%\subject{xxxxx, xxxxx}
\author{Naoki Masuda}
\email{naoki.masuda@bristol.ac.uk}
\affiliation{Department of Engineering Mathematics, University of Bristol, Clifton, Bristol, United Kingdom}

%%%% Abstract text to be placed here %%%%%%%%%%%%
\begin{abstract}
Evolutionary game theory predicts that cooperation in social dilemma games is promoted when agents are connected as a network. However, when networks are fixed over time, humans do not necessarily show enhanced mutual cooperation. Here we show that reinforcement learning (specifically, the so-called Bush-Mosteller model) approximately explains the experimentally observed network reciprocity and the lack thereof in a parameter region spanned by the benefit-to-cost ratio and the node's degree. Thus, we significantly extend previously obtained numerical results.
\end{abstract}
%%%%%%%%%%%%%%%%%%%%%%%%%%%

\maketitle

Human society is built upon cooperation among individuals. 
However, our society is full of social dilemmas, where cooperative actions, which are costly to individuals, appear to be superseded by non-cooperative, selfish actions that exploit cooperative others \cite{Dawes1980,Axelrod1984,Kollock1998}. There are several mechanisms that explain cooperative behavior in social dilemma situations \cite{Nowak2006,Sigmund2010,Rand2013a}. The evolutionary game theory has provided firm evidence that static networks enhance cooperation as compared to well-mixed populations under generous conditions, with the effect being called spatial reciprocity (in the case of finite-dimensional networks) and network reciprocity (in the case of general networks)  \cite{Nowak1992,Nowak2006,Ohtsuki2006,Szabo2007,Perc2013}. This finding is in alignment with broadly made observations that humans as well as animals are reasonably considered to interact on contact networks where a node is an individual \cite{Easley2010,Newman2010,Barabasi2016}.

However, a series of laboratory experiments using human participants involved in the prisoner's dilemma game (PDG) has produced results that are not necessarily consistent with spatial and network reciprocity. In fact, the structure of networks (e.g., scale-free, random, and lattice) did not correlate with the propensity of human cooperation in the PDG \cite{Traulsen2010,Cassar2007,Grujic2010,Rand2011,Suri2011,Gracia-Lazaro2012,Grujic2012}.  In contrast, Rand \textit{et al.} have shown that humans present
network reciprocity if the benefit-to-cost ratio, a main parameter of the PDG, is larger than the degree of nodes in the network (i.e., number of neighbors per player) \cite{Rand2014}, which is consistent with the prediction of evolutionary game theory \cite{Ohtsuki2006}.
Note that the earlier experimental studies used smaller benefit-to-cost ratio values
\cite{Traulsen2010,Cassar2007,Grujic2010,Rand2011,Suri2011,Gracia-Lazaro2012,Grujic2012}.
 
The theoretical results in Ref.~\cite{Ohtsuki2006} are derived from the probability of fixation of cooperation, i.e., the probability that a unanimity of cooperation is reached before that of defection under weak selection (i.e., the difference between the strength of cooperator and that of defector is assumed to be small). While theoretically elegant, unanimity under weak selection may be different from the population dynamics taking place in laboratory experiments with human participants, such as those in Ref.~\cite{Rand2014}. In laboratory experiments, the unanimity of cooperators is hard to be reached. The aim of the present study is to look for an alternative mechanism that explains behavioral results under the PDG on networks.
 
We hypothesize that a type of reinforcement learning implemented as a strategy of players produces game dynamics that are consistent with the aforementioned experimental results regarding network reciprocity.
In particular, aspiration-based reinforcement learning \cite{Bush1955,Rapoport1965,Macy1991,Bendor2001,Macy2002}, with which players modulate their behavior based on the magnitude of the earned reward relative to a threshold, has been successful in explaining conditional cooperation behavior and its variants called moody conditional cooperation \cite{Cimini2014,Ezaki2016}. 
Furthermore, aspiration-based reinforcement learning, not evolutionary game theory, yielded the absence of network reciprocity in numerical simulations \cite{Cimini2015}. In the present paper, we vary the benefit-to-cost ratio and the node's degree, two key parameters in the discussion of network reciprocity in the literature, to show that aspiration-based reinforcement learning gives rise to network reciprocity under the conditions consistent with the previous experimental study \cite{Rand2014}. In this way, we significantly extend the previous numerical results \cite{Cimini2015}.

%This paper is organized as follows. First, the definitions of the repeated PDG on networks, the two network treatments, and the BM model are given in Sec. \ref{sec:model}. Then, we report the results in Sec. \ref{sec:results}. 
%Finally we summarize the results and discuss their implications in Sec. \ref{sec:discussions}.

\section{Model}\label{sec:model}
\subsection{Prisoners' dilemma game on networks}
Consider players placed on nodes of a network. 
They repeatedly play the donation game, which is a special case of the PDG, over $t_{\rm{max}}$ rounds as follows. 
In each round, each player selects either to cooperate (C) or defect (D), and a donation game occurs on each edge in both directions. 
The submitted action (i.e., C or D) is consistently used toward all neighbors.
On each edge, a cooperating player pays cost $c$ to benefit the other player by $b$. 
If a player does not cooperate (i.e., D), both the focal player and the other player get nothing. We impose $b>c>0.$
For example, if both players constituting an edge cooperate, both gain $b-c$. Each player is assumed to have $k$ neighbors. Therefore, a player submitting C loses $-kc$ and gains $b$ multiplied by the number of cooperating neighbors. After the donation game has taken place bidirectionally on all edges, each player's final payoff in this round is determined as the payoff that the focal player has gained, averaged over the $k$ neighbors.

\subsection{Static- and shuffled-network treatments}
We compare the propensity of cooperation between static and dynamically shuffled networks, mimicking the situation of a laboratory experiment \cite{Rand2014}. In both static- and shuffled- network treatments, the network in each round is a ring network in which each node has $k$ neighbors, where $k$ is an even number (Fig. 1).
Each player is adjacent to $k/2$ players on each side on the ring. 
In the static-network treatment, the position of the players is fixed throughout all the rounds. 
In the shuffled-network treatment, while the network structure is fixed over rounds, we randomize the position of all the players after each round.

\subsection{BM model}

We consider players that obey the Bush-Mosteller (BM) model of reinforcement learning to update actions over rounds \cite{Bush1955,Rapoport1965,Macy1991,Macy2002}. We use the following variant of the BM model \cite{Masuda2011,Ezaki2016}. 
Each player has the intended probability of cooperation, $p_t$ ($t=1,\ldots, t_{\rm{max}}$)
as the sole internal state. Probability $p_t$ is updated in response to the payoff obtained in the previous round, denoted by $r_{t-1}$, and the previous action, denoted by $a_{t-1}$, as follows:
\begin{equation}
p_{t} = \begin{cases}
p_{t-1} + (1-p_{t-1})s_{t-1} & (a_{t-1} = \text{C}, s_{t-1}\geq 0),\\
p_{t-1} + p_{t-1} s_{t-1} & (a_{t-1} = \text{C}, s_{t-1}< 0),\\
p_{t-1} - p_{t-1} s_{t-1} & (a_{t-1} = \text{D}, s_{t-1}\geq 0),\\
p_{t-1} - (1-p_{t-1})s_{t-1} & (a_{t-1} = \text{D}, s_{t-1}< 0).
\end{cases}\label{eq:pt}
\end{equation}
In Eq.~\eqref{eq:pt} the stimulus, denoted by $s_{t-1}\in (-1,1)$, is defined by
\begin{equation}
s_{t-1} = \tanh{\left[\beta (r_{t-1} - A)\right]},
\label{eq:def stimulus}
\end{equation}
where $\beta>0$ and $A$ are the sensitivity parameter and aspiration level, respectively.
The action selected in the previous round is reinforced if the realized payoff is larger than the aspiration level, i.e., $r_{t-1}-A>0$.
Conversely, if the payoff is smaller than the aspiration level, the previous action is suppressed. 
For example, when a player submitted C in the previous round and the obtained payoff was larger than the aspiration level,
the stimulus is positive. Then, the probability of cooperation is increased in the next round [according to the first line in the RHS of Eq.~\eqref{eq:pt}]. Note that the updating scheme [Eq. \eqref{eq:pt}] guarantees $p_t\in (0,1)$ if $p_1\in (0,1).$
We set $p_1 = 0.8$, which roughly agrees with the observations made in the previous laboratory experiments \cite{Traulsen2010,Rand2011,Rand2014}.

In each round, players are assumed to misimplement the action to submit the action opposite to the intention (i.e., D if the player intends C, and C if the player intends D) with probability $\epsilon$ \cite{Nowak1993,Nowak1995,Masuda2011,Ezaki2016}.
Thus, the actual probability of cooperation is given by $\tilde{p}_t = p_t(1-\epsilon)+(1-p_t)\epsilon$.

\section{Results}\label{sec:results}
We consider two values of $b/c$, i.e., $b/c=2$ and $6$ by setting $(b,c) = (2,1)$ and $(b,c)=(6,1)$, respectively.

Numerically calculated fractions of cooperative players are compared between the two treatments in Fig.~\ref{fig:timecourse}.
When the node's degree, $k$, is small (i.e., $k=2$) and $b/c$ is large (i.e., $b/c=6$), cooperation is more frequent in the static-network treatment than the shuffled-network treatment. This result is consistent with the previous experimental results \cite{Rand2014}.
When $b/c=2$, this effect is not observed, which is also consistent with the experimental results \cite{Cassar2007,Grujic2010,Traulsen2010,Rand2011,Suri2011,Gracia-Lazaro2012,Grujic2012,Rand2014}.

To examine the robustness of the results shown in Fig.~\ref{fig:timecourse}, we carried out simulations for a region of the $A$--$\epsilon$ parameter space and four values of $k$.
We did not vary $\beta$ ($=0.2$) because $\beta$ did not considerably alter the behavior of  the players unless it took extreme values \cite{Ezaki2016}.
With $b/c=2$, the fraction of cooperative players averaged over the first $25$ rounds is shown in Figs.~\ref{fig:sweep}(a) and \ref{fig:sweep}(b) for the static and shuffled networks, respectively.  The difference between the two types of networks, shown in Fig.~\ref{fig:sweep}(c), is small in the entire parameter region, in particular for large $k$, suggesting a marginal effect of network reciprocity. 
In contrast, when $b/c=6$, the fraction of cooperators is larger in the static-network than the shuffled-network treatment in a relatively large region of the $A$--$\epsilon$ parameter space [Figs.~\ref{fig:sweep}(e), \ref{fig:sweep}(f), and \ref{fig:sweep}(g)]. As $k$ increases, the difference between the two treatments decreases. 
In summary, a static as opposed to shuffled network promotes cooperation only when $b/c$ is large and $k$ is small. 
These results are consistent with the experimental findings \cite{Rand2014}. 

Network reciprocity is attributed to assortative connectivity between cooperative players \cite{Nowak1992,Ohtsuki2006,Szabo2007,Perc2013}. In other words, cooperation can thrive if a cooperator tends to find other cooperators at the neighboring nodes.
To measure this effect, we defined the assortment by $P(\rm{C}|\rm{C}$$;t) -P(\rm{C}|\rm{D}$$;t)$, where $P(\rm{C}|\rm{C}$$;t)$ is the probability that a neighbor of a cooperative player is cooperative in round $t$, and $P(\rm{C}|\rm{D}$$;t)$ is the probability that a neighbor of a defective player is cooperative in round $t$ \cite{VanVeelen2009,Rand2014}. For various values of $A$ and $\epsilon$, the assortment values in the static-network treatment averaged over the first 25 rounds are shown in Figs. \ref{fig:sweep}(d) and \ref{fig:sweep}(h) for $b/c=2$ and $b/c=6$, respectively. 
The figures indicate that the assortment tends to be positive when cooperation is more abundant in the static-network than shuffled-network treatment regardless of the value of $b/c$, suggesting that cooperative players are clustered in these parameter regions. In the shuffled treatment, we confirmed that the assortment was $\approx 0$ in the entire parameter region.

\section{Conclusions}\label{sec:discussions}

We have numerically shown that an aspiration-based reinforcement learning model, the BM model, produces network reciprocity if and only if the benefit-to-cost ratio in the donation game is large relative to the node's degree. The results are consistent with the previous experimental findings \cite{Cassar2007,Grujic2010,Traulsen2010,Rand2011,Suri2011,Gracia-Lazaro2012,Grujic2012,Rand2014}.
In addition to network reciprocity, the BM model also accounts for the conditional cooperation, which is hard to explain by evolutionary game theory \cite{Cimini2014,Ezaki2016,Horita2017}. Aspiration-based reinforcement learning may be able to describe cooperative behavior of humans and animals in broader contexts.
Finally, we remark that, although network reciprocity is not observed in the shuffled-network treatment, in both theory and experiments, dynamic linking treatments that allow players to strategically sever and create links promote cooperation in laboratory experiments \cite{Fehl2011,Rand2011,Wang2012a,Jordan2013,Shirado2013}. Evolutionary game theory predicts cooperation under dynamic linking \cite{Zimmermann2004,Eguiluz2005,Zimmermann2005,Pacheco2006,Pacheco2006a,Gross2008,Perc2010}. Reinforcement learning may also account for enhanced cooperation under dynamic linking. 

\section*{Acknowledgments}

TE acknowledges the support provided through PRESTO, JST (No. JPMJPR16D2) and Kawarabayashi Large Graph Project, ERATO, JST. NM acknowledges the support provided through, CREST, JST (No. JPMJCR1304) and Kawarabayashi Large Graph Project, ERATO, JST.

\bibliographystyle{plos2015}
\bibliography{RL}
%\begin{thebibliography}{10}
%\end{thebibliography}

\clearpage
\begin{figure}[p]
\centering\includegraphics[width=80mm]{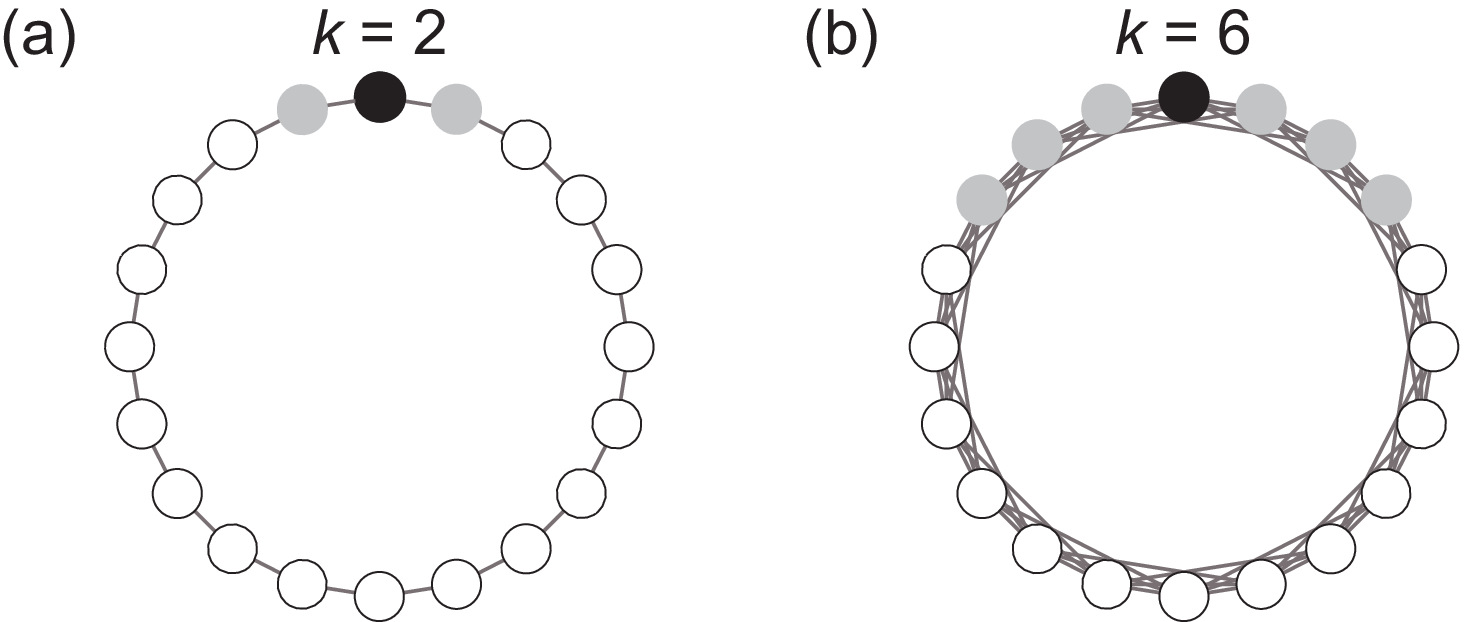}
%%% where xxxxxx name represents "figurename.eps"
\caption{Ring networks composed of $N=20$ players. The player represented by a black circle is adjacent to $k$ players represented by gray circles. (a) $k=2$. (b) $k=6$.}
\label{fig:network}
\end{figure}

\clearpage
\begin{figure*}[p]
\centering\includegraphics[width=160mm]{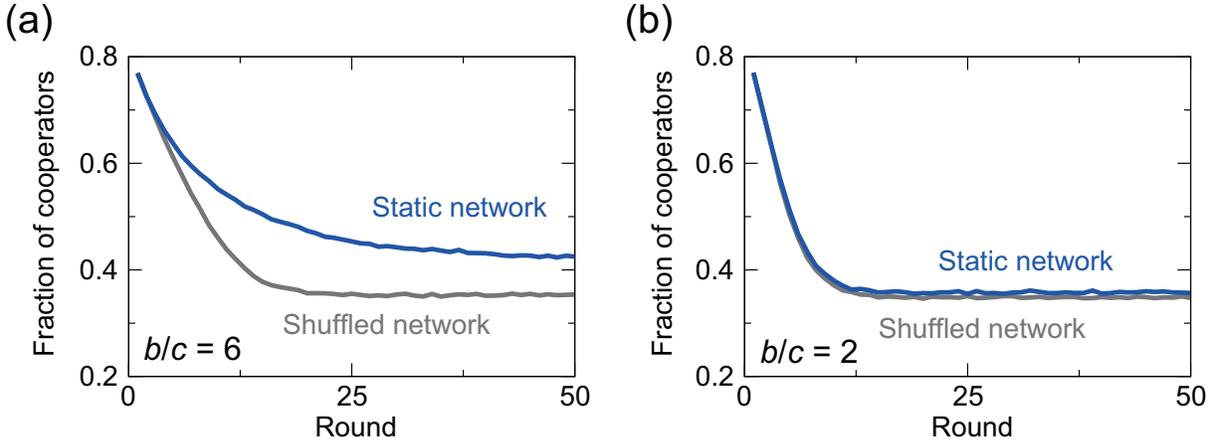}
%%% where xxxxxx name represents "figurename.eps"
\caption{Fraction of cooperative players in each round, averaged over $10^3$ simulations. We set $k=2,N=100, t_{\rm{max}}=50, \beta=0.2,A=1.0,$ and $\epsilon=0.05$. (a) $b/c =6$. (b) $b/c=2$.}
\label{fig:timecourse}
\end{figure*}

\clearpage
\begin{figure*}[p]
\centering\includegraphics[width=170mm]{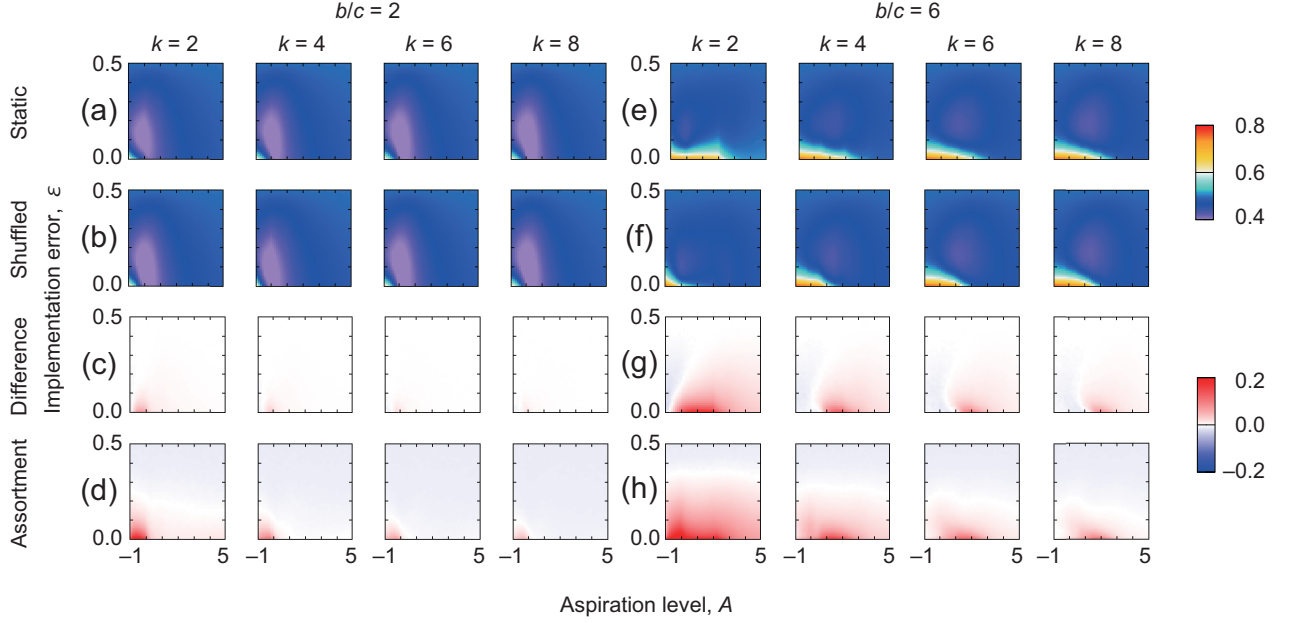}
%%% where xxxxxx name represents "figurename.eps"
\caption{Fraction of cooperative players under the static-network treatment [(a) and (e)] and the shuffled-network treatment [(b) and (f)]. The difference between the fraction of cooperation in the static and shuffled networks is shown in (c) and (g). The assortment for the static networks is shown in (d) and (h). We set $N=100$ and $\beta=0.2.$ (a)--(d) $b/c=2$. (e)--(h) $b/c=6.$
To calculate the fraction of cooperators and the assortment, we take averages over the first $25$ rounds and $10^3$ simulations.}
\label{fig:sweep}
\end{figure*}

\end{document}